

\magnification=1200

\font\twelvebf = cmbx12 scaled \magstep2
\font\bigbf=cmbx10 scaled \magstep2
\font\titlefont=cmbx10 scaled\magstep1
\font\smc=cmcsc10

\newcount\nchap         \nchap=0
\newcount\nsect
\newcount\nssec
\newcount\equanumber
\newcount\refnumber     \refnumber=0

\def\chapter #1{\relax\global\advance\nchap by 1
             \nsect=0
             \equanumber=0
             \goodbreak\bigskip\bigskip\noindent\hskip0pt
             \twelvebf\number\nchap.\  #1
             \rm\medskip\noindent\hskip0pt}

\def\section #1{\relax\global\advance\nsect by 1
             \nssec=0
             \goodbreak\bigskip\medskip\noindent\hskip0pt
             \bigbf\number\nchap.\number\nsect.\  #1
             \rm\vskip4pt\noindent\hskip0pt}

\def\parag #1{\relax\global\advance\nssec by 1
           \goodbreak\bigskip\medskip
           {\centerline{\bf\number\nchap.\number\nsect.\number\nssec.\ #1}}
           \rm\bigskip\noindent\hskip0pt}

\def\ref {\bigskip\bigskip\bigskip
            \centerline{\bf References.}\nobreak\bigskip}

\def\eqname #1{\relax\global\advance\equanumber by 1
            \xdef#1{{\rm(\number\nchap.\number\equanumber)}}#1}
\def\eqn{\eqno\eqname}

\def\refname #1{\relax\global\advance\refnumber by 1
             \xdef#1{{\rm[\number\refnumber]}}#1}
\def\ref{\refname}

\def\p{\par\noindent}
\def\v{\rho_{\scriptscriptstyle 0}}
\def\A{\vec A}
\def\pa{\partial}
\def\pax{\pa_x}
\def\J{\vec J}
\def\DR{(\delta\rho)^2}
\def\dr{\delta\rho}
\def\N{\vec\nabla}
\def\NW{\vec\nabla\wedge}
\def\E{\vec E}
\def\Eem{\E^{em}}
\def\eem{E^{em}}
\def\B{\vec B}
\def\D{\vec D}
\def\U{\vec u}
\def\gE{g_{\scriptscriptstyle E}}
\def\gB{g_{\scriptscriptstyle B}}
\def\lap{\bigtriangleup}
\def\drz{\dr_{\scriptscriptstyle 0}}

\def\e#1{\epsilon_{#1}}
\def\ds{\displaystyle}
\def\vp{\varphi}
\def\Avp{A_{\vp}^{em}}
\def\Jvp{J_{\vp}}
\def\Par{{d\over dr}}
\def\Bem{\B^{em}}
\def\bem{B^{em}}
\def\rem{\v^{em}}
\def\subarrow#1{\mathrel{\mathop{\longrightarrow}\limits_{#1}}}

\def\ce{{\cal E}}
\def\lu{\mu_1}
\def\ld{\mu_2}
\def\ra{{\rm a}}
\def\rb{{\rm b}}
\def\rc{{\rm c}}
\def\rd{{\rm d}}

\null
\vskip-2truecm
\vskip2truecm
\vskip0.5truecm
\centerline{\titlefont  PERSISTENT SUPERCURRENTS IN A PLANAR}
\smallskip
\centerline{\titlefont  NON-RELATIVISTIC CHIRAL FLUID}

\vskip3truecm
\centerline{Pietro Donatis and Roberto Iengo}
\vskip 1.0truecm
\centerline{{\it International School for Advanced Studies,
                I-34014 Trieste, Italy and}}
\centerline{{\it Istituto Nazionale di Fisica
               Nucleare, INFN, Sezione di Trieste, Trieste, Italy}}

\vskip3.0truecm
\noindent{\bf Abstract}
\vskip.3truecm
\par
\noindent
We study the possible stationary persistent supercurrents flowing on a
cylindrical sample supporting a two-dimensional charged fluid. The
internal dynamics of the fluid is obtained by means of an effective theory
in which the fluid self-interacts through a $U(1)$ gauge field. We find
that the presence of persistent supercurrents depends on what kind of gauge
field it is. In particular the current is zero if it is a Maxwell gauge
field, and it is  maximal if it is a Chern-Simons gauge field. There is an
intermediate behaviour in presence of both Maxwell and Chern-Simons term.
Therefore it appears that persistent supercurrents are possible only if the
fluid is chiral.
\vfill
\eject

\chapter{Introduction and summary.}

In this paper we study the possibility that stationary supercurrents can
flow, like in a superconductor, in a planar non-relativistic fluid which
looks incompressible in the sense that all the excitations have an energy
gap and thus there are no gapless compressional modes.

This fluid is described by the most general effective action in two space
and one time dimensions including $U(1)$ gauge fields. As it is well known
(references \ref\JACKT\ \ref\JACKD), in $2\!+\!1$ dimensions the generic
action with at most two derivatives for the gauge fields includes both a
Maxwell-like term and a parity and time reversal violating Chern-Simons
term. A gauge theory of that kind has been introduced long ago and
extensively studied also for possible relations with Chern-Simons theory
describing fractional statistics particles and rather speculative
applications to high $T_c$ superconductors (see reference \ref\WILC\ for a
general introduction).

Here we study an effective action where the $U(1)$ gauge field is coupled
to a non-relativistic matter field. This effective theory was studied for
the pure Chern-Simons case in reference \ref\NOIN. The possible vortex
excitations of this system for the Chern-Simons case and for the Maxwell
case were discussed in reference \ref\NOIP, and later also in \ref\RUSSI.
(An effective theory formally similar to the one we investigate here, in
the Chern-Simons limit, was used to describe the Fractional Quantum Hall
Effect in references \ref\ZHK, \ref\LEEF, \ref\ZHANG. Notice that the
physical system we consider is completely different from the quantum Hall
setting, in particular there is no uniform magnetic field, provided by a
given external device).

One can easily derive the spectrum (see section 2) and find that the
possible excitations have an energy gap. Thus, this fluid does not appear
similar to a standard superconductor (nor to the standard anyonic
superconductors \ref\HMRW, \ref\FHL, \ref\CWWH, \ref\WZ, \ref\MORI), which
possesses a gapless excitation that plays the role of the Higgs field when
coupled to the electromagnetic field. Nevertheless, we specifically
consider the possibility of stationary currents flowing around in an
idealized cylindrical sample, as the one of figure 1a, much like a
supercurrent in a superconducting solenoid. We consider the cylinder made
of a pile of many two dimensional annuli, on which the fluid's current
flows around. We also consider the fluid to be charged and coupled to
ordinary three-dimensional electromagnetism, and we study in particular the
magnetic field produced by the possible supercurrents.

We find in section 3 that the possibility of supercurrents depends very
much on the relative weight of the Maxwell term as compared to the
Chern-Simons term, in the two-dimensional (gauge) dynamics. An important
role is played by the electrostatic force which arises when there is a
charge fluctuation (we assume that our system, fluid plus background, is
overall electrically neutral. However we assume that the background is
rigid, and thus a fluctuation in the density of the charged fluid gives
rise to an electrostatic field). In the limit where the Maxwell term
dominates, no currents are possible in the bulk of the sample -- in this
case  the currents are concentrated at the edges and are small. Thus in
this limit our sample does not work like a superconducting solenoid.

On the contrary, in the limit where the Chern-Simons term dominates, the
current can flow freely like in a superconductor and our system behaves
like a superconducting solenoid (of the same thickness, of course), the
resulting magnetic field and flux being the same as the ones due to a
supercurrent for an ordinary superconductor. In the intermediate cases we
find intermediate behaviours.

Since there is a gap in the spectrum, the flow of the current is protected
against external disturbances, in the sense that as far as the energy
carried by the external sources is less than the gap, no interaction is
possible.

Thus, for a sizable Chern-Simons term, we find that, despite being
non-standard, our fluid has some properties similar to standard
superconductors. Actually we studied \NOIN\ already other properties of the
non-relativistic fluid coupled to a three-dimensional $U(1)$ gauge field
dominated by the Chern-Simons term and we found interesting results showing
similarities with superconductors. In particular, we studied the possible
penetration of an external magnetic field (Meissner effect) in the bulk of
an idealized sample of many two-dimensional layers where the fluid lies. We
found an intermediate behaviour, namely if the magnetic field is orthogonal
to the layers (that is, in a configuration similar to the one studied here)
the system behaves in the same way as a type II superconductor, whereas if
the the field is parallel to the layers it can penetrate much more easily,
going to zero inside with an inverse power law, rather than with an
exponential law like in the standard case.

Thus, putting together the previous results and the ones presented in this
paper, we can conclude that a chiral fluid (chirality being related to the
parity breaking Chern-Simons term) shows interesting properties which could
qualify it, in some respect, as a kind of non-standard superconductor.

\chapter{The effective lagrangian.}

Here we study a non-relativistic theory in $2\!+\!1$ dimensions described
by the following lagrangian:
$$
{\cal L}={1\over 2}\gE\left(\pa_0A_i-\pa_iA_0\right)^2-
{1\over 2}\gB\left(\e{ij}\pa_iA_j\right)^2-
$$
$$
-\alpha\e{\lambda\mu\nu}A_{\lambda}\pa_{\mu}A_{\nu}+
{1\over 2m}\phi^{\dagger}\D^2\phi+
{i\over 2}\left(\phi^{\dagger}\dot\phi-\dot\phi^{\dagger}\phi\right)+
A_0\left(|\phi|^2-\v\right)-
V(|\phi|)\ .
\eqn\lagr\
$$
Here $\phi(\vec x,t)$ is a complex field which plays the role of order
parameter and is related to the density by the relation:
$$
\rho(\vec x,t)=|\phi(\vec x,t)|^2\ .
\eqn\
$$
The covariant derivative is $D_i\!=\!\pa_i-iA_i$. Notice that $A_0$
is coupled to $\dr\!=\!\rho-\v$, and that $\v$ is the average density
representing the neutralizing background necessary for the consistency of
the theory. Since for conservation of the total number of particles the
integral over all the surface of $\dr$ is zero, we can write it as the
divergence of some vector field $\U(\vec x)$ which we choose to be
irrotational $\NW\U\!=\!0$:
$$
\dr=\N\cdot\U
\eqn\paolo
$$
Equation \lagr\ is the most general non-relativistic gauge invariant
lagrangian with minimal coupling to a vector potential, once the maximum
number of two derivatives is required. First note that beside the Maxwell
term there is also a Chern-Simons term with coupling constant $\alpha$.
Notice also that the coupling constants for the electric and the magnetic
part of the Maxwell term are different. They are equal only if the theory
is Lorentz invariant, that is considering a relativistic theory (here we
draw the reader's attention to the fact that in $2\!+\!1$ dimensions these
coupling constants have the dimension of a length and $\alpha$ is
dimensionless). In the following, for definiteness, we take
$V(|\phi|)\!=\!{1\over 2}\lambda\left(|\phi|^2-\v\right)^2$.
\p
Notice in particular that varying \lagr\ with respect to $A_i$, $A_0$ one
gets (in the gauge $\N\cdot\A\!=\!0$):
$$
\gE\pa_0(\pa_0A_i-\pa_iA_0)+\gB\e{ij}\pa_jB=
J_i-2\alpha\e{ij} (\pa_jA_0-\pa_0A_j)
\eqn\maxp
$$
$$
\gE\lap A_0=\dr-2\alpha B\ .
\eqn\maxs
$$
We have defined:
$$
\dr=\rho-\v
\qquad\qquad
B=\e{ij}\pa_iA_j
\qquad\qquad
J_i={1\over 2mi}\left(\phi^{\dagger}\pa_i\phi-\pa_i\phi^{\dagger}\phi-
2iA_i\phi^{\dagger}\phi\right)\ .
\eqn\usual
$$
Thus for $\alpha\!=\!0$ we recover the two-dimensional Maxwell equations
(though with two different couplings). For $\gE\!=\!0$, on the other hand,
from equation \maxs\ we get $\dr\!=\!2\alpha B$ which is the usual
Chern-Simons constraint relating the field strength to the matter density.
\p
Later on we will also consider the fluid to be electrically charged and
thus coupled to the true, three-dimensional electromagnetic field.

\section{The spectrum.}

Now we will study the energy spectrum of the small field perturbations of
our Maxwell-Chern-Simons theory.
\p
We take the following parameterization:
$$
\phi=fe^{i\theta}\ .
\eqn\
$$
and rewrite lagrangian \lagr\ keeping only quadratic terms in the fields.
With this choice, and in the gauge $\N\!\cdot\!\A\!=\!0$, equation
\lagr\ becomes (in this gauge one can put $A_i\!=\!\e{ij}\pa_j\psi$ and see
that $\int d^2x\ \e{ij}A_i\pa_0A_j\!=\!0)$:
$$
{\cal L}={1\over 2}\gE\left[\left(\pa_0A_i\right)^2+
\left(\pa_iA_0\right)^2\right]-
{1\over 2}\gB\left(\pa_iA_j\right)^2+
$$
$$
-2\alpha\e{ij} A_0\pa_iA_j+
{1\over 2m}\left[-{1\over 4\v}\left(\pa_i\dr\right)^2-
\v\left(\pa_i\theta\right)^2-\v A_i^2\right]+
\theta\pa_0(\dr)+A_0\dr-{1\over 2}\lambda\DR\ .
\eqn\
$$
{}From this lagrangian we get the following equations:
$$
{\delta{\cal L}\over\delta A_0}=-\gE\lap A_0-2\alpha\e{ij}\pa_iA_j+\dr=0
\eqn\eqp
$$
$$
{\delta{\cal L}\over\delta A_i}=
-\gE\pa_0^2A_i+\gB\lap A_i-2\alpha\e{ij}\pa_jA_0-{\v\over m}A_i=0
\eqn\eqs
$$
$$
{\delta{\cal L}\over\delta\theta}={\v\over m}\lap\theta+\pa_0(\dr)=0
\eqn\eqt
$$
$$
{\delta{\cal L}\over\delta(\dr)}={1\over 4m\v}\lap\dr-\pa_0\theta+A_0-
\lambda\dr=0\ .
\eqn\eqq
$$
Multiplying \eqs\ by $\e{ki}\pa_k$ we get:
$$
-\gE\pa_0^2B+\gB\lap B+2\alpha\lap A_0-{\v\over m}B=0\ .
\eqn\here
$$
Now multiplying equation \eqt\ by $\pa_0$ and equation \eqq\ by $\lap$ we
get
$$
\lap\dot\theta\!=\!-{m\over\v}\pa_0^2(\dr)\ ,
\eqn\every
$$
and
$$
{1\over 4m\v}\lap^2\dr-\lap\dot\theta+\lap A_0-\lambda\lap\dr\!=\!0\ .
\eqn\where
$$
We can now eliminate $\theta$ and $A_0$ from equations \here\ and
\where\ by using \eqp\ and \every. By taking
$B=(B_0e^{i(\ce t-\vec p\cdot\vec x)}+c.c.)$ and
$\dr=(\drz e^{i(\ce t-\vec p\cdot\vec x)}+c.c.)$ we thus obtain:
$$
\ce^2B_0-\left({4\alpha^2\over\gE^2}+{\v\over m\gE}\right)B_0+
{2\alpha\over\gE^2}\drz-{\gB\over\gE}(p_x^2+p_y^2)B_0=0
\eqn\
$$
$$
\ce^2\drz-{\v\over m\gE}\drz+{2\alpha\v\over m\gE}B_0-
{\lambda\v\over m}(p_x^2+p_y^2)\drz-{1\over 4m^2}(p_x^2+p_y^2)^2\drz=0\ .
\eqn\
$$
Therefore the values of the energy for $\vec p\!=\!0$ are the eigenvalues
of the above system:
$$
\ce^2_{1,2}={2\alpha^2\over\gE^2}+{\v\over m\gE}\pm
{2\alpha^2\over\gE^2}\sqrt{1+{\gE\v\over m\alpha^2}}\ .
\eqn\
$$
Notice that $\ce^2_{1,2}$ are both positive. If $\alpha\!\to\!0$, that is
if there is no Chern-Simons term, the two eigenvalues are equal:
$$
\ce_1=\ce_2=\sqrt{\v\over m\gE}\ .
\eqn\
$$
If, on the other hand, we take the opposite limit $\gE\!\to\!0$ then only
one of the eigenvalues remains finite whereas the other one diverges:
$$
\eqalign{
&\ce_1={2|\alpha|\over\gE}\to\infty\cr
&\ce_2={\v\over 2m|\alpha|}\ .\cr
}
\eqn\
$$
That is, in general there are two propagating degrees of freedom each with
its energy gap. In pure Maxwell theory the energy gaps are degenerate. This
degeneracy is resolved by adding the Chern-Simons term. If there is only
the Chern-Simons term the theory possesses only one  degree of freedom with
an energy gap, the other one being frozen. This value of the energy gap for
the pure Chern-Simons case has already been derived in reference \NOIN.

\chapter{Stationary currents.}

In this section we study the possibility of stationary currents. We would
like to study a configuration that, although idealized, is near to a
setting relevant for a realistic device which could carry superconducting
currents. Therefore we take the surface where our two-dimensional fluid
lies to be an annulus of radii $r_2\!>\!r_1$ such that $r_2\!-\!r_1\!=\!L$.
We consider a pile of many of these annuli separated by a spacing $d$, see
figure 1a. We assume that $L/r_1\!\ll\!1$, but still $L$ is macroscopic,
that is $L/d\!\gg\!1$ where $d$ is any length of the order of few atomic
distances such as the inter-annuli spacing. Furthermore, the total length
$L_z$ of the cylinder is supposed to be very large as compared to the radii
$r_{1,2}$. We study the possible configuration where the fluid's current
flows around in each annulus, see figure 1b, and it is uniform with respect
to $z$, that is it  is the same in each annulus. Thus our cylindrical
configuration will act as a solenoid. Later we will compute the
three-dimensional true magnetic field inside the solenoid, due to the
current.
\goodbreak
\null
\vskip7truecm
\nobreak
\centerline{Figure 1}
\vskip1truecm
\noindent
{}From the two-dimensional point of view the annulus is equivalent to a
strip which is finite in the $x$ direction, $-L/2\!\le\!x\!\le\!L/2$, and
periodic in the $y$ direction, $0\!\le\!y\!\le\!2\pi R_M$, see figure 1c
(with $R_M\!=\!{r_1+r_2\over 2}$, remember that in this approximation
$L/R_M\!\ll\!1$). We consider a possible current induced by a quantized
phase for the matter field, as for a vortex:
$$
\phi=|\phi|e^{ipy}
\eqn\vor
$$
where $p\!=\!{n\over R_M}$, $n$ integer, is fixed. The current flows around
the cylinder as in figure 1b.
\p
Notice that we consider a configuration with cylindrical symmetry,
therefore we take (in the gauge $\N\cdot\A\!=\!0$):
$$
\eqalign{
\A=(A_x,A_y)&=\bigl(0,A(x)\bigr)
\qquad
B=\pax A\cr
\U=\bigl(u(x),0\bigr)
\qquad
\dr(x)&=\pax u(x)
\qquad
|\phi|=\sqrt{\dr(x)+\v}\ .\cr
}
\eqn\
$$
As we said we assume uniformity in the $z$ direction and we consider the
limit where the cylinder's length $L_z$ is very large as compared to $R_M$,
so we can write the hamiltonian of the model as:
$$
\eqalign{
{H\over L_z}=\int dxdy\,\biggl\{
{1\over 2}\gB B^2&-{1\over 2m}\phi^{\dagger}\D^2\phi+V(|\phi|)\biggr\}+
\cr
&+{1\over 2\gE}(\dr-2\alpha B)\ast{1\over(-\lap)}\ast(\dr-2\alpha B)\ .
\cr}
\eqn\ham
$$
Here the last term comes from integrating out the auxiliary field $A_0$ and
the symbol $\ast$ means convolution.
\p
It is convenient to express this hamiltonian as the integral of a local
hamiltonian density. To this aim we rewrite the last term of \ham\ in terms
of the two-dimensional ``electrostatic'' field $\E$ determined by the
equation:
$$
\N\cdot\E=-{1\over\gE}(\dr-2\alpha B)
\eqn\bais
$$
and imposing boundary conditions such as to eliminate possible zero modes
which do not contribute to the hamiltonian \ham. This amounts to requiring
(in our configuration with cylindrical symmetry we have uniformity in the
$y$ coordinate, therefore $\E\!=\!(E(x),0)\bigr)$:
$$
E(x<-L/2)=E(x>L/2)=0\ .
\eqn\boundp
$$
The hamiltonian can then be  rewritten as
$$
{H\over L_z}=\int dxdy\,\left\{{1\over 2}\gE E^2+
{1\over 2}\gB B^2-{1\over 2m}\phi^{\dagger}\D^2\phi+V(|\phi|)\right\}
\eqn\ham
$$
We have to fix completely the boundary conditions. We assume that:
$$
A(x<-L/2)=A(x>L/2)=0
\eqn\bounds
$$
that is, we assume $A$, which represents in general a propagating excitation
of our fluid, to be zero outside the sample i.e. $\!$where there is no
fluid. Notice that, by subtracting the irrelevant constant
$c_0\!=\!-{1\over 2}\lambda\v N$ where $N$ is the fixed total particles'
number $N\!=\!\int|\phi|^2$, we can rewrite
$\int V\!=\!{1\over 2}\lambda\int|\phi|^4$ which vanishes outside the
sample.
\p
One can also say that, by definition, $\dr\!=\!0$
outside the sample. Since the fluctuation of the particles' total number in
a given domain is the flux of $\U$ through its boundary, and the particles'
total number is fixed, the boundary conditions for $u$ are
$$
u(x<-L/2)=u(x>L/2)=0
\eqn\boundt
$$
Thus, keeping into account \boundp, we see that the hamiltonian density
vanishes outside the sample, consistently with the fact that it describes
the dynamics of the fluid which is confined in the strip.
\p
We can check that the boundary condition \boundp\ is the one which
correctly ensures that the hamiltonian \ham\ implements the Chern-Simons
dynamics in the limit $\gE\!\to\!0$. In fact, the solution of equation
\bais, with the boundary conditions \boundp, \bounds\ and \boundt\ gives:
$$
E(x)=-{1\over\gE}\int\limits_{-{L\over 2}}^x dx^{\prime}\dr(x^{\prime})+
{2\alpha\over\gE}A(x)=-{1\over\gE}\bigl[u(x)-2\alpha A(x)\bigr]\ .
\eqn\
$$
Thus we see that, since finite energy requires $\gE E^2$ to be bounded, one
recovers in the limit $\gE\!\to\!0$:
$$
A(x)={1\over 2\alpha}u(x)
\eqn\
$$
which is indeed the solution of the Chern-Simons constraint
$B\!=\!{1\over 2\alpha}\dr$ with the boundary conditions of \bounds\ and
\boundt.
\p
To this hamiltonian we now add one extra term which plays an essential role
in what follows. This term represents the true, three-dimensional
electrostatic interaction between the fluctuations of the charged matter.
Of course, the whole system, fluid plus background, is overall electrically
neutral. But if the background is rigid, as we assume here, a local density
fluctuation of the charged fluid gives rise to electrostatic forces. This
yields a {\it real} electric field which obeys to the three-dimensional
Maxwell equation:
$$
\N\cdot\Eem=e\dr^{(3)}\ ,
\eqn\
$$
where $\dr^{(3)}\!=\!\dr/d$ is the three dimensional density and $e$ is the
electric charge. (for other effects of this electrostatic contribution we
refer to \NOIN). We can then compute this electrostatic contribution. The
three-dimensional Maxwell equation can be written as (in our uniform in $y$
and in $z$ configuration):
$$
\pax\eem(x)={e\over d}\dr={e\over d}\pax u\ .
\eqn\
$$
This equation can be integrated yielding:
$$
\eem={e\over d}u
\eqn\
$$
Therefore the electrostatic contribution to the hamiltonian is
$$
{d\over 2}(\Eem)^2={e^2\over 2d}u^2
\eqn\
$$
Thus, after taking everything into account, the hamiltonian of the system
can be written as:
$$
{H\over 2\pi R_ML_z}=\int dx\,\biggl\{
{1\over 2\gE}(u-2\alpha A)^2+
{1\over 2}\gB (\pax A)^2+
{1\over 2m}(p-A)^2\rho+
{1\over 2m}(\pax|\phi|)^2+
$$
$$
+{1\over 2}\lambda(\pax u)^2+
{e^2\over 2d}u^2\biggr\}
\eqn\hamiltonian
$$
Before going on it is maybe worth to stress again the difference
between $E(x)$ and $\eem(x)$. The first one is the two-dimensional
``electric field'' coming from the internal dynamics of the fluid. The
second one is the actual, real, three-dimensional electric field due to
electrostatic interaction of the charged matter.

\section{A heuristic analysis.}

We perform a heuristic analysis by considering a situation where the matter
is uniformly distributed and there is no ``magnetic field'', that is:
$$
\dr=0\qquad\qquad B=\pax A=0\ .
\eqn\
$$
This, in particular, implies $\rho\!=\!\v$ and $\pax|\phi|\!=\!0$. In this
case the variation of the hamiltonian \hamiltonian\ gives the equations
$$
\eqalign{
&{\delta\over\delta A}\left({H\over 2\pi R_ML_z}\right)=
{2\alpha\over\gE}(2\alpha A-u)+{\v\over m}(A-p)=0\cr
&{\delta\over\delta u}\left({H\over 2\pi R_ML_z}\right)=
{1\over\gE}(u-2\alpha A)+{e^2\over d}u=0\ .\cr
}
\eqn\
$$
The solution of these equations is (in this heuristic analysis we forget
the boundary conditions):
$$
\eqalign{
&A_{\ast}=\ds{\ds{\v\over m}\over\ds{\v\over m}+
\ds{4\alpha^2e^2\over d+e^2\gE}}p\cr
&u_{\ast}=\ds{2\alpha\v d\over\v(d+e^2\gE)+4m\alpha^2e^2}p\ .\cr
}
\eqn\eursol
$$
This corresponds to the current density:
$$
J={\v\over m}\ds{\ds{4\alpha^2e^2\over d+e^2\gE}\over\ds{\v\over m}+
\ds{4\alpha^2e^2\over d+e^2\gE}}p\ .
\eqn\
$$
Notice that if $e^2\!=\!0$, that is if there is no electrostatic effect,
the density current is zero.
\p
Notice also that in the Maxwell limit ($\alpha\!\to\!0$ or
$\gE\!\to\!\infty$) the current density is still zero.
\p
Therefore the current density is different from zero only if there is some
Chern-Simons amount in the fluid's dynamics. In particular for the case of
pure Chern-Simons, with $\alpha\!\to\!\infty$, we recover the maximum value
for the current:
$$
J\;\subarrow{\gE\to 0}\;{\v\over m}\,{4m\alpha^2e^2\over\v d+4m\alpha^2e^2}\,p
\;\subarrow{\alpha\to\infty}\;{\v\over m}p\ .
\eqn\
$$

\section{A more detailed analysis.}

The assumptions of the previous section are physically too restrictive
because the actual physical situation can have $\dr$ and $B$ different from
zero. Nevertheless in this section we show that a more accurate computation
leads substantially to the same results, apart from a correction which is
different from zero only in regions very close to the sample's borders.
\p
In order to keep the discussion reasonably transparent, we make the
simplificatory assumptions that $\v\!\gg\!|\dr|$ and that
$1/m\!\ll\!\lambda$. With these assumptions we can write:
$$
\rho(p-A)^2\simeq\v(p-A)^2
\qquad\qquad
{1\over 2m}(\pax |\phi|)^2={1\over 8m\rho}(\pax\dr)^2\simeq 0\ .
\eqn\simpl
$$
We stress that our results would hold also in general. In fact, the
equilibrium configuration of the fluid, far from the borders, will be the
one of section 3.1. What happens in details near to the borders will depend
on the details of the hamiltonian, and we have chosen to present the
simplified discussion based on \simpl.
\p
In this way the field equations now read:
$$
\eqalign{
&{\delta\over\delta A}\left({H\over 2\pi R_ML_z}\right)=
-\pax^2A+{2\alpha\over\gE\gB}(2\alpha A-u)+{\v\over m\gB}(A-p)=0\cr
&{\delta\over\delta u}\left({H\over 2\pi R_ML_z}\right)=
-\pax^2u+{1\over\lambda\gE}(u-2\alpha A)+{e^2\over\lambda d}u=0\ .\cr
}
\eqn\
$$
These equations can be rewritten in matricial form as follows:
$$
(-{\bf 1}\pax^2+M)\Psi=\Phi
\eqn\matr
$$
where ${\bf 1}$ is the unit two by two matrix, and
$$
M=\left(\matrix{\ra&\rb\cr \rc&\rd\cr}\right)=
\left(\matrix{
\ds{4\alpha^2\over\gE\gB}+\ds{\v\over m\gB}&
-\ds{2\alpha\over\gE\gB}\cr
-\ds{2\alpha\over\gE\lambda}&
\ds{1\over\gE\lambda}+\ds{e^2\over\lambda d}\cr
}\right)
\qquad
\Psi=\left(\matrix{
A\cr
u\cr
}\right)
\qquad
\Phi=\left(\matrix{
\ds{\v\over\gB m}p\cr
0\cr
}\right)\ .
\eqn\
$$
Note that $\det M\!\ne\!0$.
\p
The general solution of equation \matr\ can be written as
$$
\Psi(x)=\Psi_0+\Psi_{\ast}
\eqn\
$$
where
$$
\Psi_{\ast}\!=\!M^{-1}\Phi=\left(\matrix{A_{\ast}\cr u_{\ast}\cr}\right)
\eqn\
$$
is one particular solution and $A_{\ast}$ and $u_{\ast}$ are given by
equation \eursol. $\Psi_0$ is the general solution of the homogeneous
equation $(-{\bf 1}\pax^2+M)\Psi\!=\!0$. The boundary conditions discussed
in section 3 are $\Psi(-L/2)\!=\!\Psi(L/2)\!=\!0$. Therefore we can write:
$$
\left(\matrix{A(x)\cr u(x)\cr}\right)=
{\cosh(\lu x)\over\cosh(\lu L/2)}
\left(\matrix{A_1\cr u_1\cr}\right)+
{\cosh(\ld x)\over\cosh(\ld L/2)}
\left(\matrix{A_2\cr u_2\cr}\right)+
\left(\matrix{A_{\ast}\cr u_{\ast}\cr}\right)
\eqn\sol
$$
where
$$
\mu^2_{1,2}={1\over 2}\left(\ra+\rd\pm\sqrt{(\ra-\rd)^2+4\rb\rc}\right)
\eqn\
$$
are the eigenvalues of the matrix $M$, and
$$
u_{1,2}\!=\!{\mu^2_{1,2}-\ra\over\rb}A_{1,2}\ .
\eqn\
$$
One can verify that $\mu^2_{1,2}$ are real positive. Let us mention their
values in some limiting cases. For the case of pure Maxwell, i.e.
$\alpha\!=\!0$, the fields $A(x)$ and $u(x)$ decouple and the two
eigenvalues are
$$
\mu_1^2\!=\!{\v\over\gB m} \;\; \hbox{(relevant for $A$)}
\qquad
\mu_2^2\!=\!{1\over\lambda\gE}\!+\!{e^2\over\lambda d} \;\;
\hbox{(relevant for $u$)}\ .
\eqn\
$$
In the opposite limit $\gE\!\to\!0$ the relevant field configuration
is $u\!=\!2\alpha A$, and this is reflected in the eigenvalues which
turn out to be
$$
\mu_1^2\!\to\!\infty
\qquad
\mu_2^2\!=\!\ds{\ds{\v\over m}+
\ds{4e^2\alpha^2\over d}\over\gB+4\alpha^2\lambda}\ .
\eqn\
$$
Imposing the boundary conditions, we determine $A_{1,2}$ by:
$$
\left(\matrix{
1&1\cr
\ds{\lu^2-\ra\over\rb}&\ds{\ld^2-\ra\over\rb}\cr}\right)
\left(\matrix{A_1\cr A_2\cr}\right)=-
\left(\matrix{A_{\ast}\cr u_{\ast}\cr}\right)\ .
\eqn\
$$
Notice that the two eigenvalues can be equal only for
${\alpha\over\gE}\!=\!0$ and
${\v\over m\gB}\!=\!{1\over\gE\lambda}+{e^2\over\lambda d}$ in which case
$M$ is proportional to the identity, and in \sol\ we can write
$A_1\!=\!A_2$ and $u_1\!=\!u_2$ with no relation between $A_1$ and $u_1$.
The boundary conditions give in this case $A_1\!=\!-A_{\ast}$ and
$u_1\!=\!-u_{\ast}$.
\p
Since $\mu^2_{1,2}$ are expressed in terms of the microscopic parameters of
the hamiltonian, we have $L\!\cdot\!\mu_{1,2}\!\gg\!1$. Thus,
$A(x)\!=\!A_{\ast}$ and $u(x)\!=\!u_{\ast}$ apart from very near to the
edges, that is for $|x|\!\simeq\!L/2$, and therefore we recover the
solution found in the previous heuristic analysis.

\section{Magnetic field.}

In this subsection we compute the actual magnetic field present inside the
cylinder (see figure 1a) due to the supercurrent computed above.
\p
It is clear that we have to take now a three-dimensional view-point. Our
three-dimensional system is now the cylinder embedded in the
three-dimensional space. Therefore we will use the cylindrical coordinates
$(r,\vp,z)$ as in figure 1a. We remind that we consider $L\!\ll\!r_{1,2}$
so that the cylinder can be thought of zero thickness and we take its
radius to be \hbox{$R_M\!=\!{r_1+r_2\over 2}$}. The three-dimensional
density of the fluid can be written in this approximation as
$\rho^{(3)}\!=\!{\v\over d}L\cdot\delta (r-R_M)$.
\p
Furthermore we suppose $L_z\!\gg\!R_M$ so that it is possible to keep far
from the cylinder's edges and therefore to disregard the edge effects.
\p
With these definitions the three-dimensional current density flowing all
round the cylinder is:
$$
\Jvp={e\v L\over md}\left[p-e\Avp(r)-A\right]\delta(r-R_M)\ .
\eqn\
$$
Here $\Avp(r)$ is the electromagnetic vector potential describing the
magnetic field produced inside the cylinder. Since we limit ourselves to
the region far from the edges of the cylinder, we can assume $\Avp$
independent of $z$.
\p
$A$ is the (two-dimensional) vector potential describing the fluid's
dynamics on the strip considered in the previous section.
\p
We can reproduce all the computation of the previous section taking into
account for $\Avp$ simply with the substitution $p\!\to\!p-e\Avp$. Therefore
we now have, instead of \eursol:
$$
A=\ds{\ds{\v\over m}\over\ds{\v\over m}+\ds{4\alpha^2e^2\over d+e^2\gE}}
(p-e\Avp)
\eqn\
$$
We have neglected the deformation of $A$ very near to the annuli's edges, as
discussed in the previous section.
Now we use this expression for $A$ to solve Maxwell's equation
$\NW\Bem\!=\!\J$ inside the cylinder (see reference \ref\LECH). The solution
of this equation is:
$$
\Avp={e\rem\over m}\, p\, {1\over\ds 1+\ds{e^2\rem R_M\over 2m}}
\left[{r\over 2}\Theta(R_M-r)+{R_M^2\over 2r}\Theta(r-R_M)\right]\ .
\eqn\
$$
where:
$$
\rem=\ds{\ds{4e^2\alpha^2\over d+e^2\gE}\over\ds{\v\over m}+
\ds{4e^2\alpha^2\over d+e^2\gE}}\cdot\v {L\over d}\ .
\eqn\
$$
Here $\Theta$ is the step function. This yields the current density:
$$
\Jvp={e\rem\over m}{p\over\ds 1+\ds{e^2\rem R_M\over 2m}}\delta(r-R_M)\ ,
\eqn\pucci
$$
We can also compute the magnetic field inside the cylinder using the
cylindrical coordinate relation $\bem_z\!=\!{1\over r}\Par(r\Avp)$:
$$
\bem_z={e\rem\over m}{p\over\ds 1+\ds{e^2\rem R_M\over 2m}}\Theta(R_M-r)\ .
\eqn\
$$
This result is the same as the one of a solenoid in which the current
density \pucci\ flows. We can also compute the flux of the magnetic field in
the cylinder:
$$
\Phi(\bem)=\pi R_M^2\bem_z=
\pi R_M^2{e\rem\over m}{1\over\ds 1+\ds{e^2\rem R_M\over 2m}}p\ .
\eqn\
$$
Remembering that $p\!=\!{n\over R_M}$, we see that for
$e\rem R_M^2\!\to\!\infty$ we get
$$
\Phi(\bem)\longrightarrow{2n\pi\over e}\ .
\eqn\
$$
That is, the total flux is quantized in the above limit, as expected for a
vortex-like current.
\p
We stress that, as we said, the result for the magnetic field and flux is
exactly the same as it would have been obtained for an ordinary
superconductor flowing on a cylindrical surafce, $\rem$ being its surface
density. (It is a general fact for any superconducting current flowing on a
cylindrical surface, that the flux quantization is strictly speaking
obtained just in the above limit).
\p
Note that $\rem$ vanishes in the Maxwell limit $\gE\!\to\!\infty$ (or
$\alpha\!=\!0$) and $\rem\!=\!\v {L\over d}$ for $\alpha\!\to\!\infty$.
Therefore we see again that, in order that our idealized device could work
as a superconducting solenoid, there cannot be only a Maxwell dynamics, but
the chiral Chern-Simons term must play an essential role.
\bigskip
\bigskip
\bigskip
\noindent
$\underline{\hbox{Acknowledgements.}}$ One of us, R. I., would like to thank
Adam Schwimmer for useful discussions on the content of section 2.1.
\bigskip
\bigskip
\bigskip
\centerline{\bf References.}
\nobreak
\bigskip
\noindent
\JACKT\ {\smc R. Jackiw, S. Templeton},
{\it Phys. Rev.} {\bf D23} (1981), 2291.
\vskip 0.3truecm

\noindent
\JACKD\ {\smc S. Deser, R. Jackiw, S. Templeton},
{\it Ann. Phys.} {\bf 140} (1982), 372.
\vskip 0.3truecm

\noindent
\WILC\ {\it Fractional Statistics and Anyon Superconductivity},
edited by F. Wilczek, World Scientific, (1990).
\vskip 0.3truecm

\noindent
\NOIN\ {\smc P. Donatis, R. Iengo},
{\it Nucl. Phys.} {\bf B415}[FS] (1994), 630 .
\vskip 0.3truecm

\noindent
\NOIP\ {\smc P. Donatis, R. Iengo},
{\it Phys. Lett.} {\bf B320} (1994), 64.
\vskip 0.3truecm

\noindent
\RUSSI\ {\smc I.V. Barashenkov, A.O. Harin},
{\it Phys. Rev. Lett.} {\bf 72} (1994), 1575.
\vskip 0.3truecm

\noindent
\ZHK\ {\smc S.C. Zhang, T.H. Hansson, S. Kivelson},
{\it Phys. Rev. Lett.} {\bf 62} (1989), 82.
\vskip 0.3truecm

\noindent
\LEEF\ {\smc D.H. Lee, M.P.A. Fisher},
{\it Int. J. Mod. Phys.} {\bf B5} (1991), 2675.
\vskip 0.3truecm

\noindent
\ZHANG\ {\smc S.C. Zhang},
{\it Int. J. Mod. Phys.} {\bf B6} (1992), 25.
\vskip 0.3truecm

\noindent
\HMRW\ {\smc B.I. Halperin, J. March-Russel, F. Wilczek},
{\it Phys. Rev.} {\bf B40} (1989), 8726.
\vskip 0.3truecm

\noindent
\FHL\ {\smc A.L. Fetter, C.B. Hanna, R.B. Laughlin},
{\it Phys. Rev.} {\bf B39} (1989), 9679.
\vskip 0.3truecm

\noindent
\CWWH\ {\smc Y.H. Chen, F. Wilczek, E.Witten, B.I. Halperin},
{\it Int. J. Mod. Phys.} {\bf B3} (1989), 1001.
\vskip 0.3truecm

\noindent
\WZ\ {\smc X.G. Wen, A. Zee},
{\it Phys. Rev.} {\bf B41} (1990), 342.
\vskip 0.3truecm

\noindent
\MORI\ {\smc H. Mori},
{\it Phys. Rev.} {\bf B42} (1990), 184.
\vskip 0.3truecm

\noindent
\LECH\ {\smc R. Iengo, K. Lechner},
{\it Nucl. Phys.} {\bf B346} (1990), 551.
\vskip 0.3truecm
\vfill
\eject

\centerline{\bf Figure caption}
\bigskip
\noindent
Figure 1.
\medskip
\noindent
\item{a)} The cylinder as a pile of many annuli
\item{b)} A single annulus
\item{c)} The annulus as a periodic strip
\vfill
\eject
\bye